\documentclass[11pt]{article}
\usepackage{graphicx,amssymb,amsmath,times,cite}
\begin{document}
\begin{titlepage}
\vspace{4cm}

\begin{center}{\Large \bf On a suggestion relating topological and quantum mechanical entanglements}\\
\vspace{1cm} M. Asoudeh\footnote{email:asoudeh@mehr.sharif.edu}, \hspace{0.5cm} V. Karimipour
\footnote{email:vahid@sharif.edu}, \hspace{0.5cm} L. Memarzadeh\footnote{email:laleh@mehr.sharif.edu},\\
\vspace{0.5
cm} A. T. Rezakhani \footnote{email:tayefehr@mehr.sharif.edu}\\
\vspace{1cm} Department of Physics, Sharif University of Technology,\\
P.O. Box 11365-9161,\\ Tehran, Iran
\end{center}
\vskip 3cm

\begin{abstract}
We analyze a recent suggestion \cite{kauffman1,kauffman2} on a
possible relation between topological and quantum mechanical
entanglements.  We show that a one to one correspondence does not
exist, neither between topologically linked diagrams and entangled
states, nor between braid operators and quantum entanglers. We
also add a new dimension to the question of entangling properties
of unitary operators in general.
\end{abstract}
\end{titlepage}

\begin{figure}
\setlength{\unitlength}{1mm}
 \centering
\includegraphics[width=3.5cm,height=2.8cm]{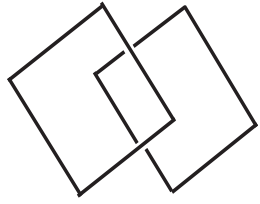}
\caption{A knot representation of the disentangled state
$|\phi\rangle \otimes |\psi\rangle$.} \label{unlinkedcircles}
\end{figure}

\begin{figure}
\setlength{\unitlength}{1mm}
 \centering
\includegraphics[width=3.5cm,height=2.8cm]{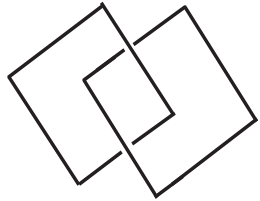}
\caption{A knot representation of a bipartite entangled state.}
\label{linkedcircles}
\end{figure}
\section{Introduction}
\begin{figure}[tp]
\setlength{\unitlength}{1mm}
 \centering
\includegraphics[width=4cm,height=4cm]{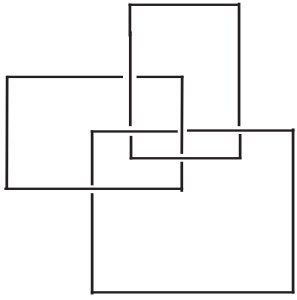}
\caption{ A knot representation of the {\small{GHZ}} state
$\frac{1}{\sqrt{2}}(|0,0,0\rangle+|1,1,1\rangle)$, known also as
Borromean rings.} \label{ghzfigure}
\end{figure}

In a recent series of papers \cite{kauffmanref, kauffman1, kauffman2},
it has been argued that there may be a relation between quantum mechanical entanglement and topological
entanglement. This hope has been raised by some formal similarities between entanglement of quantum mechanical
states which is an algebraic concept and linking of closed curves which is a topological concept. Let us begin by
simple definitions of these two
concepts and the basic idea of a correspondence put forward in the above papers.\\
A pure quantum state of a composite system $AB$ (a vector
$|\Psi\rangle $ in the tensor product of two Hilbert spaces
${\cal{H}}_A\otimes {\cal{H}}_B$) is called entangled if it can
not be written as a product of two vectors , i.e. $ |\Psi\rangle
\ne |\psi\rangle_A \otimes |\phi\rangle_B $. The simplest
entangled pure states occur when ${\cal{H}}_A$ and $ {\cal{H}}_B$
are two dimensional with basis vectors $ |0\rangle $ and $
|1\rangle$, called a qubit in quantum computation literature. For
brevity in the following we will not write the subscripts $A$ and
$B$ explicitly. A general state of two qubits
\begin{equation}\label{psi}
  |\Psi\rangle = a|0,0\rangle +b|0,1\rangle +c|1,0\rangle +d|1,1\rangle
\end{equation}
is entangled provided $ ad-bc\ne 0$.\\
On the other hand two curves can be in an unlinked position like
the one shown in figure (\ref{unlinkedcircles}) or in a linked
position like the one shown in figure (\ref{linkedcircles}). One
is tempted to view the two unlinked curves as a topological
representation of a disentangled quantum state and the two linked
curves as a
representation of an entangled state.\\
In the same way that cutting any of the curves in figure
(\ref{linkedcircles}) removes the topological entanglement,
measuring one of the qubits of the state $|\Psi\rangle $ in
(\ref{psi}) in any basis (not
necessarily the $\{|0\rangle,|1\rangle\}$ basis), disentangles the quantum state.\\
More evidence in favor of this analogy is provided by figure
\ref{ghzfigure} \cite{kauffman2}, which provides an alleged
topological equivalent for the so-called {\small GHZ} state
\cite{ghz}
\begin{eqnarray}\label{ghz}
&  |{\rm{GHZ}}\rangle := \frac{1}{\sqrt{2}}(|0,0,0\rangle +
  |1,1,1\rangle).
\end{eqnarray}

\begin{figure}[tp]
\setlength{\unitlength}{1mm}
 \centering
\includegraphics[width=4cm,height=4cm]{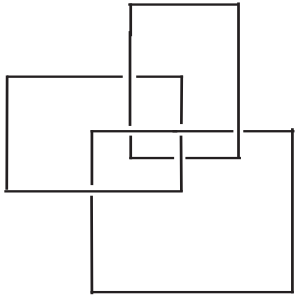}
\caption{ A knot representation of the entangled state
$|\Phi\rangle =\frac{1}{2}(|0,0,0\rangle + |1,1,0\rangle +
  |1,0,1\rangle + |0,1,1\rangle) $.} \label{Wfigure}
\end{figure}

In this figure cutting any of the three curves, leaves the other
two curves in an unlinked position, in the same way that measuring
any of the three subsystems in the {\small GHZ} state in the
$\{|0\rangle, |1\rangle\} $ basis,
leaves the other two subsystems in a disentangled state. \\
One may be tempted to make a general correspondence between
topologically linked diagrams and entangled states or vice versa.
For example while figure \ref{ghzfigure} corresponds to the
{\small GHZ} state, a slight modification of the crossings of this
link diagram, as shown in figure \ref{Wfigure}, may correspond to
the following state
\begin{eqnarray}\label{W}
  &|\Phi\rangle = \frac{1}{2}(|0,0,0\rangle + |1,1,0\rangle +
  |1,0,1\rangle + |0,1,1\rangle).
\end{eqnarray}

If one measures one of the subsystems in this state in the
$\{|0\rangle, |1\rangle\}$  basis, the other two subsystems are
projected onto an entangled state, in the same way that cutting
out any of the component curves in figure \ref{Wfigure} leaves the
other two components in a linked position.\\

A natural question arises as to how serious and deep such a
correspondence may be. Certainly such a relation, if exists, will
be much fruitful for both fields and it is worthwhile to explore
further the possibility of its existence. We should stress here
that we only want to analyze one particular suggestion
\cite{kauffman1, kauffman2}, regarding a possible correspondence
between topological and quantum mechanical entanglement. We are
not concerned here with other aspects of the relation between
topology and quantum computation or quantum mechanics.  These
avenues of study have been followed in \cite{freedman1,
freedman2,kitaev, freedman3} where the possibility of doing fault
tolerant quantum computation by using topological degrees of
freedom of certain systems with anyonic excitations or the design
of quantum algorithms for calculating topological invariants of
knots are analyzed.
\\

 It is the aim of this paper to shed more light on these analogies
and to study more closely the similarities and differences
between these the above types of entanglement. The overall
picture that we obtain is that these analogies do not point to a
deep relation between these concepts, since despite some
superficial similarities, there are many serious differences
which lead to the conclusion that such a correspondence can not
be taken seriously.

Here we list some of these differences.\\

\begin{figure}[tp]
\setlength{\unitlength}{1mm}
 \centering
\includegraphics[width=3cm,height=3cm]{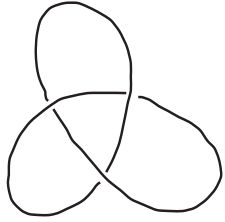}
\caption{ The trefoil knot, an example of a single component knot
which is entangled.} \label{trefoil}
\end{figure}

1- If we want to correspond any component of a linked knot with a
state of a vector space in a tensor product space, (the number of
vector spaces being equal to the number of components of the link
diagram), then we are faced with the obvious question of ''What
kind of state corresponds to a knot which is highly linked with
itself". We can imagine many topologically different one component
knots and yet we have to correspond them all to a single state in
a vector space which necessarily has no self-entanglement. Figure
(\ref{trefoil}) shows such a knot
known as trefoil knot.\\
One way out is to consider only linked diagrams whose individual
components have no self linking and to take into account the
linking between different components. But there is no natural way
to separate the linking of a component with itself from that with
others. A component may be topologically trivial by itself (when
one removes all the other components), but can not be deformed
continuously to a trivial knot due to the presence of other
components.

2- The second problem is that quantum entanglement should not
change by local unitary operations which is equivalent to local
change of basis. Therefore for such a correspondence to be valid,
two quantum mechanical states which are related to each other by
local unitary operations should correspond to topologically
equivalent diagrams. Let us see if this is the case. Consider the
examples given above: The two states (\ref{ghz}) and (\ref{W}) are
related to each other by the local action of three Hadamard
matrices ($ H :={\frac{1}{\sqrt{2}}\left(\begin{array}{cc}
  1 & 1 \\
  1 & -1
\end{array}\right)}$):
\begin{equation}\label{hadamard}
  |{\rm {GHZ}}\rangle = H\otimes H \otimes H |\Phi\rangle\hskip 1cm
  |\Phi\rangle = H\otimes H \otimes H |{\rm{GHZ}}\rangle
\end{equation}
and yet they correspond to completely inequivalent diagrams, shown
in figures \ref{ghzfigure} and
\ref{Wfigure}, respectively.\\

3- The third problem concerns the alleged relation between
''measurement" of a quantum state on one hand and ''cutting a
line" in a knot diagram on the other hand. This relation is very
questionable. The only evidence is that in some simple cases as
those mentioned above, it appears that measurement (reduction) of
a state $|\Psi\rangle $ which corresponds to a knot $K$, produces
a state $|\Psi'\rangle $ which corresponds to a knot $K'$ obtained
by cutting one of the lines of $K$. However this correspondence is
too superficial since the reduction of a wave function depends on
what value we obtain for our observable while cutting a line is an
action with a unique and predetermined result. To see this more
explicitly consider a state like
\begin{eqnarray}\label{werner}
  &|W\rangle = \frac{1}{\sqrt{3}}(|1,0,0\rangle+ |0,1,0\rangle+|0,0,1\rangle).
\end{eqnarray}
If we measure the first qubit in the computational basis $\{ |0\rangle , |1\rangle \}$ and obtain the value $0$,
the other two qubits are projected onto an entangled state, while if we obtain the value $1$, the other two
qubits are projected onto a disentangled state. Therefore one can not identify a measurement with a simple
cutting of a line
in a knot diagram. The result of the measurement also determines if the remaining state is entangled or not. \\
These examples provide sufficient reasons to abandon the kind of correspondence mentioned above. But the question
of a possible relation remains open and there may be an alternative and more tractable framework for studying
it.\\ It is well known that all knots and links can be obtained from closure of braids, the latter having a direct
relation with operators acting on tensor product spaces. Therefore it may be possible to find a correspondence
between entangling operators on the quantum mechanical side and braid operators which produce topological
entanglement on the other side. It is in order to present a short review of braid group and braid operators.

\subsection{A review of braid group}\label{2}
\begin{figure}[tp]
\setlength{\unitlength}{1mm} \centering
\includegraphics[width=4cm,height=3cm]{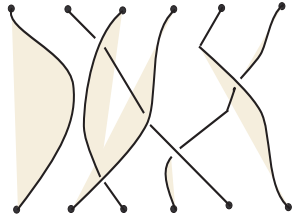}
\caption{ A braid on 6 strands, an element of the group $ B_6$.}
\label{braid}
\end{figure}

\begin{figure}[tp]
\setlength{\unitlength}{1mm} \centering
\includegraphics[width=4cm,height=3cm]{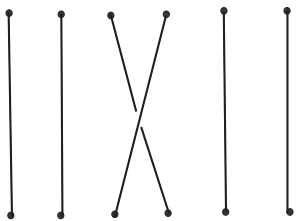}
\caption{ The generator $\sigma_3 $ of the braid group $B_6$.}
\label{sigmas}
\end{figure}

A braid on $n$ strands (figure \ref{braid}) is the equivalence
class of a collection of continuous curves joining $n$ points in a
plane to $n$ similar points on a plane on top of it. The curves
should not intersect each other but can wind around each other
arbitrarily.
Two collections of curves which can be continuously deformed to
each other are considered equivalent. There is a well-known
theorem stating that each knot can be constructed from the closure
of a braid (see \cite{knots} for a review). By closure of a braid
we mean joining the points on the lower plane to those on the
upper one by
continuous lines which lie outside all the curves of the braid.\\
The collection of all braids can be equipped with a group
structure by defining the product of two braids $ \alpha$ and $
\beta $ as the equivalence class of a braid obtained by inserting
the braid $\beta $ on top of the braid $ \alpha $. The unit
element of this group is simply the equivalence class of paths
which do not wind around each other when they go from the lower
plane to the upper one.\\ This group, called the braid group on
$n$ strands and denoted by $ B_n $, is generated by the simple
braids $ \sigma_i $, $ i = 1, \ldots, n-1$, shown in figure
\ref{sigmas}, where each $\sigma_i$ intertwines, only once, the
strands $ i $ and $ i+1 $ ( $\sigma_{i}^{-1}$ intertwines the
strands in the opposite direction). Such elements generate the
whole braid group when supplemented with the following relations
which express topological equivalence of braids as the reader can
verify:
\begin{eqnarray}\label{generators}
\sigma_i \sigma_{i+1}\sigma_i &=& \sigma_{i+1} \sigma_{i}\sigma_{i+1}\cr \sigma_{i}\sigma_{j} &=& \sigma_j
\sigma_{i}\ \ \ \ {\rm{if}} \ \ \  |i-j|\geq 2.
\end{eqnarray}
The expression of braids as elements of a group shows how to find the inverse of braids as topological objects.
For example the inverse of $\alpha = \sigma_1 \sigma_2 $ is $ \alpha^{-1} = \sigma_2^{-1} \sigma_1^{-1} $.\\ One
can obtain a representation of the braid group $B_n $ for any $n$ on the tensor product space $ V^{\otimes n}$, if
one can find a solution of the following equation in $ V^{\otimes 3}$ called hereafter the braid relation
\begin{equation}\label{braidrelation}
  (R\otimes I)(I\otimes R)(R\otimes I) =(I\otimes R)(R\otimes I)(I\otimes
  R),
\end{equation}
in which $ R: V\otimes V\rightarrow V\otimes V $ is a linear operator called a braid operator and $I$ is the
identity operator. Once such a solution is found, representations of generators of the braid group and hence the
whole braid group is obtained as follows:
\begin{eqnarray}\label{representation}
  &\sigma_i = I^{\otimes^{i-1}}\otimes R \otimes I^{\otimes^{n-i-1}}
\end{eqnarray}
where for simplicity we have used the same notation for $\sigma_i
$ and its representation.\\
Thus if we have a braid operator $R$, we can produce
representations of all kinds of braids with all the variety of
their topological entanglement. \\
Once a representation is in hand one can try to construct
invariants of knots by defining suitable traces on the space
$V^{\otimes n}$ \cite{turaev}. \\  We have now set the stage for
asking the question of a possible relation between topological and
quantum mechanical entanglement
in an appropriate way. We can ask the following questions:\\

1- Does every braid operator which produces topological
entanglement, also necessarily produces quantum entanglement? or conversely\\

2- Does every quantum entangler (an operator which entangles
product states) necessarily produces topological
entanglement?, that is , is any quantum entangler related somehow to a solution of the braid group relation?\\

We think that the answer to these questions will shed light on the question of relation between
topological and quantum mechanical entanglements.\\
We choose to investigate these questions for two dimensional spaces, since in two dimensions we have both a classification of
solutions of the braid group relation and a great deal of information about measures of quantum entanglement. \\
In the rest of this paper we try to answer the above questions and
draw our conclusions which are mainly negative, that is we
conclude that the two types of entanglement may not be related to
each other in such a direct way. This however does not exclude the
possibility that quantum computation may someday be used for
calculating topological invariants of knots
\cite{freedman1,freedman2, kitaev}.
\\
The structure of this paper is as follows. In section 2 we present
all the unitary solutions of the braid group relation in two
dimensions (4$\times$4 unitary braid operators $R$). In section 3
we collect the necessary tools for the analysis of entanglement of
states and entangling properties of operators. In section 4 we use
these tools to characterize the braid operators. Finally we end
the paper with a discussion which encompasses a summary of our
results.

\section{All unitary braid operators in two dimensions}\label{sec2}

Let $V$ be a vector space and let $ \hat{R}: V\otimes V\rightarrow
V \otimes V $ be a linear operator. The
 following equation which is a
relation between operators acting on $ V\otimes V \otimes V $ is
called the Yang-Baxter relation first formulated in studies on
integrable models \cite{Baxter}
\begin{eqnarray}\label{Yangbaxter}
&\hat{R}_{12}\hat{R}_{13}\hat{R}_{23}=\hat{R}_{23}\hat{R}_{13}\hat{R}_{12},
\end{eqnarray}
where the indices indicate on which of the three spaces, the
operator is acting non-trivially. Any solution $\hat{R}$ of the
Yang-Baxter equation provides a braid operator $R$ by the simple
relation $ R = P\hat{R}$, where $P$ is the permutation or
{\small{SWAP}} operator defined as $ P|i,j\rangle = |j, i\rangle
$. When the vector space $ V $ is two dimensional, the solutions
of Yang-Baxter equation have been classified up to the symmetries
allowed by the equation \cite{hietarinta, hlavaty, dye}. From
these solutions
 we can select those solutions of the
braid group equation which are unitary.  We should stress that
this restriction can be relaxed and one can also consider
non-unitary solutions of braid group. The reason for our interest
in unitary operators in this paper is that in quantum mechanics we
want to use these operators as quantum gates.\\
There are only two types of unitary solutions. A single one
designated as
\begin{eqnarray}
&\hskip 1cm  R=\frac{1}{\sqrt{2}}\left(\begin{array}{cccc}
  1 &  &  & 1 \\
   &  1 &-1 &  \\
   & 1 & 1 &  \\
   -1&  &  & 1
\end{array}\right)
\label{R}
\end{eqnarray}
and a continuous family of solutions
\begin{eqnarray}
 R' = \left(\begin{array}{cccc}
  a &  &  &  \\
   &  & b &  \\
   & c &  &  \\
   &  &  & d
\end{array}\right),
\label{R'}
\end{eqnarray}
where the complex parameters $ a, b, c, $ and $ d $ are pure phases, i.e. $ |a| = |b|=|c|=|d| = 1 $.\\
Note that the {\small{SWAP}} operator (denoted by $P$) is a
special kind of the matrix $R'$ for which $ a = b = c = d =1$.
For general value of its parameters, it is simply the
{\small{SWAP}} operator times a diagonal matrix. The second of
these solutions can be generalized to arbitrary dimensions, in
the form $ R'_{ij,kl} = M_{ij} \delta_{il}\delta_{jk}$, where $
|M_{ij}|=1$. We do not know of any generalizations of
the other solution.\\

\section{Entanglement of pure states and entangling power of unitary operators}

Consider a pure state of two qubits $A$ and $B$:
\begin{equation}\label{purestate}
  |\Psi\rangle_{AB} = \alpha|0,0\rangle + \beta|0,1\rangle+\gamma|1,0\rangle + \delta |1,1\rangle.
\end{equation}
The single parameter
\begin{equation}\label{concurrence}  C:=2|\alpha \delta - \beta \gamma|
\end{equation}
called the concurrence, characterizes the entanglement of this
state \cite{hillwootters,wootters}. For a product state $
|\psi\rangle \otimes |\psi'\rangle \equiv (x|0\rangle + y
  |1\rangle)\otimes (x'|0\rangle + y'|1\rangle),
$ this parameter is zero and for a maximally entangled state like
one of the Bell states $(\frac{|0,0\rangle \pm
|1,1\rangle}{\sqrt{2}} ,\frac{|0,1\rangle \pm
|1,0\rangle}{\sqrt{2}})$, it takes its maximum value of 1. We note
that the concurrence can be written as $C=|\langle \Psi^*|\sigma_y
\otimes \sigma_y |\Psi\rangle| $ where $\sigma_y$ is the second
Pauli matrix and $*$ denotes complex conjugation in the
computational basis. This also shows that the concurrence is
invariant under local transformations $|\Psi\rangle \rightarrow
U\otimes V |\Psi\rangle $,
since $U^T\sigma_yU = \sigma_y$.\\
Any other measure of entanglement, like the von Neumann entropy of
the reduced density matrices $ \rho_A$ or $\rho_B$  defined as $$
E_v(\rho):= - {\rm tr}(\rho \ln \rho)$$
 or the linear entropy defined as $$ E_l(\rho):= 1-
{\rm tr}(\rho^2)$$ can be expressed in terms of this parameter. A
simple calculation shows that the eigenvalues of the reduced
density matrix for the state in (\ref{purestate}) are
\begin{eqnarray}\label{eigenvlaues1}
&  \lambda_{\pm} = \frac{1}{2}(1\pm \sqrt{1-C^2}),
\end{eqnarray}
from which the simple expression $ E_l = \frac{1}{2}C^2 $ is
obtained for the linear entropy. \\ The concurrence, the linear
entropy or the von Neumann entropy are increasing functions of
each other, all of them vanish for a product state and take their
maximum values of $1$, $\frac{1}{2}$ and $ 1 $ respectively for
maximally entangled states. One can use any
of these measures for the characterization of entanglement of a pure state of two qubits. \\
So much for the entanglement properties of states, we now turn to the entangling properties of operators acting
on the space of two qubits.\\ The space of unitary operators acting on two qubits (the group U(4)) when viewed in
terms of entangling properties has a rich structure. Those in the subgroup U(2)$\otimes$U(2) are called local
operators. Elements of this subgroup can not produce entangled states when acting on product states. The
complement of this subgroup forms the set of non-local operators. In the set of non-local operators, those which
can produce a maximally entangled state when acting on a suitable product state are called perfect entangler
\cite{Zhang}. An example in this class is the {\small{CNOT}} operator defined as ${\rm{CNOT}}|i,j\rangle = |i,
i+j~({\rm{mod}}\ 2)\rangle $. Those non-local operators which do not have this property are called non-perfect
entanglers. Note also that there are non-local operators which can not produce any entanglement at all. An
example is the {\small{SWAP}} operator $P$ which is
incidentally a braid group operator.\\
An important concept is the local equivalence of two operators.
Let two operators $U$ and $U'$ in U(4) be related as follows:
\begin{equation}\label{local}
  U'=(k\otimes l)U(k'\otimes l'),
  \end{equation}
 where  the local operators $k, l,
k' , l' \in {\rm U(2)}$. Two such operators should be regarded
equivalent as far as their entangling properties are
concerned. \\
As far as entangling properties are concerned one may extend this
notion of equivalence to the case where the two operators are
related by the {\small{SWAP}} operator $P$, that is when $U'=UP$
or $U'=PU$, or both, since the {\small{SWAP}} operator does not
change the entanglement of a state. However the {\small{SWAP}}
operator is non-local which means that it can not be implemented
by local unitary operations on the two states. Moreover as far as
topological properties are concerned, the {\small{SWAP}} operator
is a braid operator and totally changes the topological class of a
braid. For this reason we restrict ourselves to the notion of
bi-local
equivalence as in (\ref{local}).\\

 How can we find if two such operators are equivalent? This question has been studied by many authors
 \cite{Grassl,Linden1,Linden2,Makhlin,Zhang}. The orbits of states under bi-local \cite{Makhlin,Zhang}
  and multi-local unitaries (in the case of multi-particle states) \cite{Linden1,Linden2}
 have been characterized by certain invariants. Here we use the invariants found in \cite{Makhlin,Zhang}.
 Let us define the matrix $Q$ as follows:
\begin{eqnarray}
 & Q = \frac{1}{\sqrt{2}}\left(
\begin{array}{cccc}
  1 & 0 & 0 & i \\
  0 & i & 1 & 0 \\
  0 & i & -1 & 0 \\
  1 & 0 & 0 & -i
\end{array}
  \right).
\label{Q}
\end{eqnarray}
For any matrix $U\in {\rm U(4)}$ define the following matrix:
\begin{eqnarray}
 & m(U):= (Q^{\dagger}UQ)^T(Q^{\dagger}UQ)
\label{m}
\end{eqnarray}
where $T$ denotes the transpose. Note that $Q^{\dagger}UQ$ is
nothing but the matrix expression of the operator $U$ in the Bell
basis modulo some phases. It is shown in \cite{Makhlin,Zhang} that
the followings are invariant under bi-local unitary operations:

\begin{eqnarray}\label{invariants}
&&G_1 ~=~ \frac{{\rm tr}^2[m(U)]}{16 \det U}\\
&&G_2 ~=~ \frac{{\rm tr}^2[m(U)]-{\rm tr}[m^2(U)]}{4 \det U}.
\end{eqnarray}

\subsection{Perfect entanglers}
By a perfect entangler we mean an operator which can produce
maximally entangled states when acting on a suitable product
state.  The following theorem \cite{Zhang} determines when a given
operator $U\in {\rm U(4)}$ is
a perfect entangler.\\

\textbf{Theorem \cite{Zhang}:} An operator $U\in {\rm U(4)}$ is a
perfect entangler if and only if the convex hull of
the eigenvalues of the matrix $m(U)$ contains zero.\\
We remind the reader that the convex hull of $N$ points $ p_1, p_2, \cdots, p_N $ in $\texttt{R}^n$ is the set
\begin{eqnarray}
 & {\cal C}:=\{ \sum_{j=1} \lambda_j p_j | \lambda_j \geq 0 \  \ {\rm{and}} \ \  \sum_{j=1}^N
\lambda_j = 1 \}. \label{convex hull}
\end{eqnarray}
The above criterion divides the set of non-local operators into
perfect entanglers and non-perfect entanglers.  A more
quantitative measure has been introduced in \cite{Zanardi} which
defines the entangling power of an
 operator $U$, as
\begin{equation}\label{e_p}
  e_p(U):= \overline{E(U|\phi\rangle \otimes |\psi\rangle)},
\end{equation}
where $E$ is any measure of entanglement of states and the average
is taken over all product states. To guarantee that the entangling
power of equivalent operators are equal, as it should be, the
measure of integration
is taken to be invariant under local unitary operations.\\

Equipped with the above tools we can now repose the questions raised in the introduction and ask what are the
status of the braid group solutions in the space of all operators acting on two qubits. Which of them is a
perfect entangler? If yes, are they both equivalent to some well-known perfect entanglers like {\small{CNOT}} or
else, they belong to different equivalence classes of perfect entanglers? In answering these questions we have
found some new features
of entangling properties of operators as we will discuss in the sequel. \\

\section{Entangling properties of braid operators}

In this section we want to study the entangling property of the
braid operators (\ref{R},\ref{R'}). Before proceeding we note a
point without any calculation. The {\small{SWAP}} operator is a
braid operator, (it is equal to $R'$ when $a=b=c=d=1$) and yet it
can not entangle product states. In fact $ P(|\phi\rangle \otimes
|\psi\rangle) = |\psi\rangle \otimes |\phi\rangle $. On the other
hand the operator {\small{CNOT}} is not a solution of braid group
relation and yet it is a perfect entangler. In fact when acting on
the product states
\begin{eqnarray}\label{productcnot}
&\frac{1}{\sqrt{2}}(|0\rangle + |1\rangle) \otimes |0\rangle , \hskip 1cm \frac{1}{\sqrt{2}}(|0\rangle +
|1\rangle) \otimes |1\rangle \cr &\frac{1}{\sqrt{2}}(|0\rangle - |1\rangle)\otimes |0\rangle , \hskip 1cm
\frac{1}{\sqrt{2}}(|0\rangle - |1\rangle)\otimes |1\rangle
\end{eqnarray}
it produces the maximally entangled Bell states
\begin{eqnarray}\label{bellcnot}
&\frac{|0,0\rangle + |1,1\rangle}{\sqrt{2}}, \hskip 1cm
\frac{|0,1\rangle + |1,0\rangle}{\sqrt{2}} \cr &\frac{|0,0\rangle
- |1,1\rangle}{\sqrt{2}}, \hskip 1cm \frac{|0,1\rangle -
|1,0\rangle}{\sqrt{2}}.
\end{eqnarray}
However by this example we do not want to rush to the conclusion that there is absolutely no relation between
braid operators and quantum mechanical entangling operators. The reason is that although the operator
{\small{CNOT}} may not be a braid operator itself, it may be locally equivalent to a braid operator via bi-local
unitary
operators.\\
Therefore to study the entangling properties of braid operators we
have to extract their non-local properties which is achieved by
first calculating their invariants.  For comparison we note that
the invariants of {\small{CNOT}} turn
out to be  $G_1 = 0, $ and $ G_2 = 1$.\\

\textbf{1:} For the braid operator $R$ we have:
\begin{eqnarray}\label{m(R)}
  &m(R) = -i \left(\begin{array}{cccc}
     &  &  & 1 \\
     &  & 1 &  \\
     & 1 &  &  \\
    1 &  &  &  \
  \end{array}
  \right)
\end{eqnarray}
from which we obtain the invariants
\begin{equation}\label{G(R)}
  G_1(R) = 0 \hskip 1cm  G_2(R) = 1.
\end{equation}
These invariants are the same as the invariants of {\small{CNOT}},
and hence this braid operator is equivalent to a quantum
mechanical perfect entangler. It is readily seen from (\ref{R})
that when acting on the computational
basis $\{|0,0\rangle, |0,1\rangle, |1,0\rangle , |1,1\rangle\}$ it produces the Bell basis. \\

\textbf{2:} For the continuous family $R'$ we obtain after simple
calculations
\begin{eqnarray}\label{m(R')}
  m(R') = {\rm{diag}} (ad, bc, bc, ad)
\end{eqnarray}
This leads to the invariants
\begin{eqnarray}\label{G(R')}
  G_1(R') &=& \frac{-(ad+bc)^2}{4abcd} \equiv -\frac{(1+\Delta)^2}{4\Delta}\\
  G_2(R') &=& -1+2G_1,
\end{eqnarray}
where $\Delta:=\frac{ad}{bc}$. \\
The last relation $G_2 = -1 + 2 G_1 $ shows that none of the members of this family is equivalent to
{\small{CNOT}}. In fact they are not equivalent to any controlled operator $U_c$ (such an operator acts on the
second qubit as $U$ only if the first qubit called the control qubit is in the state $|1\rangle $, otherwise it
acts as a unit operator). In fact a simple calculation shows that for all such controlled operators we have

\begin{equation}\label{invariantsofcontrolledU}
 G_2(U_c) = 1+2G_1(U_c).
\end{equation}
which means that even if the first invariant of such an operator
is made equal to that of $R'$, their second invariants can not be
equal to each other and thus under no condition the braid operator
$R'$ can be locally
equivalent to a controlled operator $U_c$.\\

Now that none of these braid operators are equivalent to {\small{CNOT}}, is there any perfect entangler among
them?\\

To answer this question we note that the eigenvalues of the matrix $ m(R')$ are $ ad $ and $bc$. The convex hull
of these points in the complex plane is a line which passes through the origin only if the parameter $\Delta$ is
real. Since all the parameters $ a, b, c $ and $ d $ are of unit modulus, this parameter can only have two
values, namely $\pm 1 $. The value $\Delta=1$ should be excluded, since in that case the eigenvalues are all
equal and the convex hull degenerates to a point. Thus the braid operators $R'$ are perfect entanglers only if
$\Delta = -1$. Since this same parameter determines the invariants of $R'$, there is only one single perfect
entangler in this class up to local equivalence. We take this perfect entangler to be the following matrix with
invariants $G_1 = 0 $ and $ G_2 = -1$:
\begin{eqnarray}\label{R0}
  &R'_0:=\left(\begin{array}{cccc}
    1 &  &  &  \\
     &  & 1 &  \\
     & -1 &  &  \\
     &  &  & 1 \
  \end{array}\right).
\end{eqnarray}
It produces maximally entangled states when acting on an
appropriate product basis:
\begin{eqnarray}\label{R0product}
  &R'_0|x+\rangle |x+\rangle = \frac{1}{2}(|0,0\rangle + |0,1\rangle - |1,0\rangle+|1,1\rangle)\cr
  &R'_0|x+\rangle |x-\rangle = \frac{1}{2}(|0,0\rangle + |0,1\rangle + |1,0\rangle-|1,1\rangle)\cr
  &R'_0|x-\rangle |x+\rangle = \frac{1}{2}(|0,0\rangle - |0,1\rangle - |1,0\rangle-|1,1\rangle)\cr
  &R'_0|x-\rangle |x-\rangle = \frac{1}{2}(|0,0\rangle - |0,1\rangle + |1,0\rangle+|1,1\rangle)
\end{eqnarray}
where $|x\pm\rangle = \frac{|0\rangle \pm |1\rangle}{\sqrt{2}}$. \\
Incidentally we note that the operator $R'$ when acting on the
above product basis produces an orthonormal basis of states all
with the same value of concurrence $ C = \frac{|1-\Delta|}{2}$,
\begin{eqnarray}\label{Rproduct}
  &R'|x+\rangle |x+\rangle = \frac{1}{2}(a|0,0\rangle +b |0,1\rangle +c |1,0\rangle+d|1,1\rangle)\cr
  &R'|x+\rangle |x-\rangle = \frac{1}{2}(a|0,0\rangle +b |0,1\rangle -c |1,0\rangle-d|1,1\rangle)\cr
  &R'|x-\rangle |x+\rangle = \frac{1}{2}(a|0,0\rangle -b |0,1\rangle +c |1,0\rangle-d|1,1\rangle)\cr
  &R'|x-\rangle |x-\rangle = \frac{1}{2}(a|0,0\rangle -b |0,1\rangle -c |1,0\rangle+d|1,1\rangle).
\end{eqnarray}

Up to now we have found that the two braid group families (the
single and the continuous one) each encompass a perfect entangler.
This finding is certainly in favor of a relation between
topological and quantum mechanical entanglements. Meanwhile we
have found another maximally entangled basis which is not
bi-locally equivalent to the Bell basis in the sense that no local
unitary can turn one into the other, since if they were this would
mean that the nonlocal operators $R'_0$ and {\small{CNOT}} which
generate them from product bases were locally equivalent which we
know is not the case. We should add that all maximally entangled
bases are equivalent to the Bell basis up to phases. This applies
also to the above basis. However these phases can be removed only
by nonlocal operations. \\
We are now faced with the following question:\\
Are there perfect entanglers which are not locally equivalent to the braid group operators?\\
To answer this question we should search for nonlocal operators
$U$ which although have different local invariants from $(G_1 = 0
, G_2=1)$ and $(G_1 = 0, G_2 = -1)$, the eigenvalues of their
$m(U)$ matrix, encompass the origin, so that they become perfect
entanglers.\\ One such matrix is the square root of the
{\small{SWAP}} operator \cite{Zhang}
\begin{eqnarray}\label{sqrtp}
  &\sqrt{P}= \left(\begin{array}{cccc}
    1 &  &  &  \\
     & \frac{1+i}{2} & \frac{1-i}{2} &  \\
     & \frac{1-i}{2} & \frac{1+i}{2} &  \\
     &  &  & 1 \
  \end{array} \right),
\end{eqnarray}
for which we have $m(\sqrt{P}) = {\rm diag} (1, 1, -1, 1)$, $ G_1
= \frac{i}{4} $ and $ G_2 = 0 $.
 This operator is a perfect entangler and can turn a suitable product state like $|x+\rangle |x-\rangle$ into
a maximally entangled state like $\frac{1}{2}(|0,0\rangle -i |0,1\rangle +|1,0\rangle -|1,1\rangle) $.
 However, there is an important difference. Unlike {\small{CNOT}} and $R'_0$ it can not maximally
entangle an orthonormal product basis. We can prove this as
follows. The most general form of an orthonormal product basis is
as follows:
\begin{eqnarray}\label{1}
|\psi_1\rangle &=& (a|0\rangle + b|1\rangle)\otimes( c|0\rangle + d|1\rangle)\cr
 |\psi_2\rangle &=& ( \overline{b}|0\rangle -\overline{a}|1\rangle)\otimes( c|0\rangle + d|1\rangle)\cr
|\psi_3\rangle &=&( e|0\rangle + f|1\rangle)\otimes( \overline{d}|0\rangle - \overline{c}|1\rangle)\cr
 |\psi_4\rangle &=&( \overline{f}|0\rangle - \overline{e}|1\rangle)\otimes( \overline{d}|0\rangle -
 \overline{c}|1\rangle),
 \end{eqnarray}
where $|a|^2+|b|^2 = |c|^2+|d|^2=|e|^2+|f|^2=1$. We now act on one
of these states, say the first one, by the operator $\sqrt{P}$ and
obtain:
\begin{eqnarray}\label{nogotheorem}
 \sqrt{P}|\psi_1\rangle &=& ac|0,0\rangle + \frac{1}{2}((1+i)ad+(1-i)bc)|0,1\rangle \nonumber\\& &+
  \frac{1}{2}((1-i)ad+(1+i)bc)|1,0\rangle + bd|1,1\rangle.\hskip 3mm
\end{eqnarray}
Such a state is maximally entangled if its concurrence is equal to
1. The concurrence is easily calculated to be
$C(\sqrt{P}|\psi_1\rangle) = |ad-bc|^2 $. Thus for this operator
to turn these orthonormal states into maximally entangled states
the following equations should be satisfied simultaneously:

\begin{eqnarray}\label{contradiction}
  |ad-bc|^2 &=&1 \hskip 1cm |\overline{b}d+\overline{a}c|^2 = 1\cr
  |cf-de|^2 &=&1 \hskip 1cm |\overline{c}e+\overline{d}f|^2 = 1.
  \end{eqnarray}
However the first two equalities when added together side by side give $(|a|^2+|b|^2)(|c|^2+|d|^2) = 2
$ which is impossible sine the left hand side is equal to 1 due to the normalization of states. This is also true for the second pair of equalities.\\
Therefore the operator $\sqrt{P}$ can not maximally entangle a
product basis. Note that although we have arrived at a
contradiction by only considering the pair of equalities obtained
from the states $|\psi_1\rangle $ and $|\psi_2\rangle $ it is not
true to conclude that this operator can not maximally entangle any
two orthonormal states. For we could have taken two orthonormal
product states as
\begin{eqnarray}\label{22}
|\phi_1\rangle &=& ( a|0\rangle + b|1\rangle)\otimes( c|0\rangle + d|1\rangle)\cr
 |\phi_2\rangle &=& ( \overline{b}|0\rangle -\overline{a}|1\rangle)\otimes( \overline{d}|0\rangle -\overline{c}|1\rangle)
 \end{eqnarray}
without running into any contradiction, i.e. the operator $\sqrt{P}$ maximally entangles the two orthonormal
product states $|x+\rangle |x-\rangle$ and $|x-\rangle |x+\rangle$.\\
This raises the hope that the braid operators may be the only perfect entanglers which have the important
property of maximally entangling a basis. This could be a substantial evidence for the existence of a relation
between topological and quantum mechanical entanglement. However we have found other classes of perfect
entanglers, locally inequivalent to the braid operators which have the above mentioned property. Each member of
the following one parameter family of operators
\begin{eqnarray}\label{U}
  &U_{\phi} = e^{-i\frac{\pi}{4}\sigma_y\otimes \sigma_y - i \phi \sigma_z \otimes \sigma_z} = \frac{1}{\sqrt{2}}\left(\begin{array}{cccc}
    e^{-i\phi} &  &  & ie^{-i\phi} \\
     &e^{i\phi} & -ie^{i\phi} &  \\
     & -ie^{i\phi} & e^{i\phi} &  \\
    ie^{-i\phi} &  &  & e^{-i\phi} \
  \end{array}\right)
\end{eqnarray}
has local invariants $G_1 = 0, G_2 = \cos 4 \phi$ and maximally
entangles the product basis $\{|0,0\rangle,$ $|0,1\rangle,
|1,0\rangle, |1,1\rangle \}$ as follows:
\begin{eqnarray}\label{basisforU}
   |0,0\rangle &\rightarrow & \frac{e^{-i\phi}}{\sqrt{2}}(|0,0\rangle + i |1,1\rangle)\cr
   |0,1\rangle &\rightarrow& \frac{e^{i\phi}}{\sqrt{2}}(|0,1\rangle -i |1,0\rangle)\cr
   |1,0\rangle &\rightarrow & \frac{e^{i\phi}}{\sqrt{2}}(-i|0,1\rangle +  |1,0\rangle)\cr
   |1,1\rangle &\rightarrow& \frac{e^{-i\phi}}{\sqrt{2}}(i|0,0\rangle + |1,1\rangle).
\end{eqnarray}
Note that although the phases $ e^{\pm i\phi} $ enter in the entangled basis states as overall phases,
nevertheless this phase is important when acting on linear combination of states and can not be removed by local
operations. \\
In view of this, we may
conclude that the braid operators have no special status among perfect entanglers.\\

We conclude this section by calculating the entangling power of
the the braid operators $R$ and $R'_0$. We use the linear entropy
$E_l(\Psi)$ for our measure of entanglement of a pure state
$|\Psi\rangle $, since calculation of the resulting integrals is
easier. This is indeed the measure used in \cite{Zanardi} for
defining entangling power of operators. As mentioned in the
introduction $E_l = \frac{1}{2}C^2 $ where $C$ is the concurrence
of the state. Thus the entangling power of an operator $U$ denoted
by $e_p(U)$ is calculated as follows: We take a product state
$|\psi\rangle | \psi'\rangle$, where $ |\psi\rangle =
\left(\begin{array}{c}
  \cos \frac{\theta}{2}e^{-i\frac{\phi}{2}} \\
\sin \frac{\theta}{2}e^{i\frac{\phi}{2}}
\end{array}\right)$ and $ |\psi'\rangle =
\left(\begin{array}{c}
  \cos \frac{\theta'}{2}e^{-i\frac{\phi'}{2}} \\
\sin \frac{\theta'}{2}e^{i\frac{\phi'}{2}}
\end{array}\right)$,
 determine the concurrence $C(U|\psi\rangle |\psi'\rangle)$ from (\ref{concurrence})
 and then calculate the following integral
\begin{eqnarray}\label{epp}
  &e_p(U) = \frac{1}{(4\pi)^2}\int \frac{C^2(U|\psi\rangle |\psi'\rangle)}{2}\sin \theta d\theta d\phi \sin \theta' d\theta'
  d\phi'.
\end{eqnarray}
We expect the following relations to hold and indeed they turn out to be correct:
\begin{eqnarray}\label{ep}
&e_p(U_{\phi}) = e_p(R'_0)=e_p({\rm{CNOT}}) \hskip 1cm
e_p(\sqrt{P})< e_p({\rm{CNOT}}).
\end{eqnarray}
Note that the operators {\small{CNOT}}, $R'_0 $ and $U_{\phi}$ are
not locally equivalent. The reason for their equal entangling
power is that they are related by the {\small{SWAP}} operator. The
second inequality is expected since the
operator $\sqrt{P}$ although a perfect entangler, can not entangle orthonormal bases. \\
Straightforward calculations along the lines mentioned above give the following explicit values:

\begin{eqnarray}\label{epps}
 &e_p(U_{\phi})= e_p(R) = e_p({\rm{CNOT}}) = e_p(R'_0) = \frac{2}{9},
\end{eqnarray}
and
\begin{eqnarray}\label{eppps}
  &e_p(R') = \frac{|ad-bc|^2}{18}\equiv \frac{|1-\Delta|^2}{18}\hskip 2cm e_p(\sqrt{P}) = \frac{1}{6}.
\end{eqnarray}

\section{Discussion}\label{sec4}
Following a suggestion by Kauffman and Lomonaco \cite{kauffman1}
we have tried to see if there is any relation between topological
and quantum mechanical entanglements. We have searched for a
possible relation from two different points of view. The first
point of view which is based on a possible correspondence between
linked knots and entangled states is easily refuted by various
counterexamples and arguments. The second viewpoint which is based
on a correspondence of braid operators and quantum mechanical
entanglement is more promising. In two dimensional spaces there is
a complete classification of braid operators. There is a
continuous family and a discrete one. We have shown that the
discrete solution is a quantum mechanical perfect entangler and
the continuous family encompasses a quantum mechanical perfect
entangler. Both these operators have the important property that
they can maximally entangle a full orthonormal basis of the space,
a property which is shared by
well-known quantum entanglers like {\small{CNOT}} but not by all of them. \\
However we have found other operators having this property and yet
not locally equivalent to the braid operators which shows that
even in this view point one can not ascribe a very special
status to the braid operators. \\
In our study we have come across new ideas and questions about
entangled states and entanglement which are outside the scope of
the title of our paper. For example we have shown that not every
perfect entangler is perfect. By this we mean that although it can
maximally entangle some product states, it may fail to do the same
for a product basis. Questions like ''How many inequivalent
classes of maximally entangled bases exist for a space $V\otimes
V$? " or ''How many inequivalent classes of perfect entanglers
exist which can maximally entangle a product basis? " have been
new to us. We hope that these questions are also new and
interesting for others.
\section*{Acknowledgement}
The authors would like to thank one of the referees for his very
critical reading of the manuscript and his very valuable comments.


\begin{thebibliography}{99}
\bibitem{kauffmanref} P. K. Aravind, Borromean entanglement of the
{\small GHZ} state, {\it{ Potentiality, Entanglement and
Passion-at-a-Distance}}, edited by R. S. Cohen, {\it{et. al.}} (
Kluwer, Dordrecht, 1997), pp. 53-59.
\bibitem{kauffman1} L. H. Kauffman and S. J. Lomonaco Jr., New Jour. Phys. {\bf{4}},
73.1 (2002).
\bibitem{kauffman2} L. H. Kauffman and S. J. Lomonaco Jr., Entanglement
criteria - quantum and topological, quant-ph/0304091.
\bibitem{ghz} D. M. Greenberger, M. Horne, and A. Zeilinger,
{\it{Bell's Theorem, Quantum Theory, and Conceptions of the Universe}}, edited by M. Kafatos (Kluwer, Dordrecht,
1989).
\bibitem{freedman1} M. H. Freedman, A. Kitaev, and Z. Wang, Comm.
Math. Phys. {\bf 227}, 587 (2002).
\bibitem{freedman2} M. H. Freedman, M. J. Larsen, and Z. Wang, Comm.
Math. Phys. {\bf 227}, 605 (2002).
\bibitem{kitaev} M. H. Freedman, A. Kitaev, M. J. Larsen and W. Wang, Bull. Amer. Math. Soc. {\bf 40}, 31 (2002).
\bibitem{freedman3} A. Kitaev, Ann. Phys. {\bf 303}, 2 (2003).
\bibitem{knots}L. H. Kauffman, {\it Knots and Physics} (World Scientific, Singapore, 2001).
\bibitem{turaev} V. Turaev, {\it Quantum Invariants of Knots and 3-Manifolds} (de Gruyter Studies in
Mathematics, 18, Walter de Gruyter and Co., Berlin, 1994).

\bibitem{Baxter}R. J. Baxter, {\it Exactly Solved Models in Statistical Mechanics} (Academic Press, London,
1982).
\bibitem{hietarinta} J. Hietarinta, J. Math. Phys. {\bf{34}}, 1725 (1993). 
\bibitem{hlavaty} L. Hlavaty, J. Phys. A: Math. Gen. {\bf{25}}, L63 (1992).
\bibitem{dye} H. Dye, Unitary solutions to the Yang-Baxter equation in
dimension four, Quant. Info. Proc. {\bf 2}, 117 (2003).
\bibitem{hillwootters} S. Hill and W. K. Wootters, Phys. Rev. Lett. {\bf{78}}, 5022
(1997).
\bibitem{wootters} W. K. Wootters, Phys. Rev. Lett. {\bf 80}, 2245
(1998).
\bibitem{Zhang} J. Zhang, J. Vala, S. Sastry, and K. B. Whaley, Phys.
Rev. A {\bf{67}}, 042313 (2003).
\bibitem{Grassl}M. Grassl, M. R\"{o}tteler, and T. Beth, Phys. Rev. A {\bf 58},
1853 (1998).
\bibitem{Linden1}N. Linden and S. Popescu, Fortschr. Phys. {\bf
46}, 567 (1998).
\bibitem{Linden2}N. Linden, S. Popescu, and A. Sudbery, Phys. Rev. Lett.
{\bf 83}, 243 (1999).
\bibitem{Makhlin}Yu. Makhlin, Quant. Info. Proc. {\bf 1}, 243
(2002).
\bibitem{Zanardi} P. Zanardi, C. Zalka, and L. Faoro, Phys. Rev. A {\bf{62}},
030301(R) (2000).


\end{thebibliography}
\end{document}